\documentclass[preprint]{revtex4-1}
\usepackage[dvips]{graphicx}
\usepackage{amsmath}
\usepackage{bm}
\usepackage{ascmac}
\usepackage{setspace}
\usepackage{mathptmx}
\usepackage[version=3]{mhchem}

\usepackage{here}

\usepackage{color}

\usepackage[version=3]{mhchem}
\usepackage{here}

\begin{document}
\title{Effects of attractive inter-particle interaction on cross-transport coefficient between mass and heat in binary fluids }
\author{Tatsuma Oishi$^1$,Yuya Doi$^1$,Yuichi Masubuchi$^1$ and Takashi Uneyama$^{1*}$\thanks{*Corresponding author \\Email: uneyama@mp.pse.nagoya-u.ac.jp}}
\affiliation{\textsuperscript{1} Department of Materials Physics, Graduate School of Engineering, Nagoya University, Furo-cho, Chikusa, Nagoya 464-8603, Japan}

\begin{abstract}		
In some binary fluids, mass transport is observed under a temperature gradient. 
This phenomenon is called the Soret effect.
In this study, we discuss the influence of inter-particle interaction. 
We considered equimolar binary Lennard-Jones fluids with a mass contrast, whereas the interaction was common for all the particle pairs with various cut-off lengths. 
We performed molecular dynamics simulations of such fluids under equilibrium to obtain the cross-transport coefficients $L_{1q}$ between the fluxes of mass and heat. 
The simulation revealed that this quantity strongly depends on the cut-off length. 
Further, we decomposed the heat flux into kinetic and potential contributions and calculated the cross-correlations between decomposed fluxes and the mass flux. 
The result indicates that the potential contribution dominates $L_{1q}$, implying that the Soret coefficient is altered by the inter-particle interaction.
\end{abstract}
\maketitle
\section{Introduction}
A concentration gradient is induced by a temperature gradient for some binary mixtures of fluids due to the mass transfer\cite{platten}.
This phenomenon is the so-called thermodiffusion or the Ludwig-Soret effect being characterized by the Soret coefficient $S_T$ defined as follows.
\begin{equation}
 S_T=\frac{D'}{D}\label{Soret_coefficient}.
\end{equation}
Here, $D$ is the mutual diffusion coefficient and $D'$ is the thermal diffusion coefficient.
One of the fluids moves to the cold side when $S_T$ is positive, whereas it moves to the hot side when $S_T$ is negative.

{
For binary gases, the mechanism of the Soret effect has been theoretically clarified.
Chapman and Enskog\cite{chapman} theoretically described $S_T$ for dilute gas mixtures according to the rigid body collisions.
The theory has been experimentally verified for \ce{H_2} and \ce{CO_2}, and \ce{H_2} and \ce{SO_2} mixtures by Chapman and Dootson\cite{Dootson}, Bl\"{u}h et al.\cite{bluh}, and Ibbs et al.\cite{Ibbs}
}

In contrast, for liquid mixtures, the mechanism of the phenomenon has not been fully clarified yet, even for simple systems due to the strong correlation between the constitutions\cite{wirtz}.
The earliest experimental report was made for various salt solutions\cite{snowdon1960}, and studies for the other systems followed.
For instance, Tanner\cite{Tanner,Tanner1}, Prigogine et al.\cite{Prigogine.1952,Prigogine.1950} and Saxton et al.\cite{Saxton} reported the results for organic liquids.
Korsching et al.\cite{korsching} performed a series of experiments on isotope separation.
Various theoretical models have been proposed to explain these experimental data of the Soret coefficients.
Thermodynamic and phenomenological 
models have been developed. 
In such phenomenological models\cite{Denbigh,Hartmann,Hartmann2,Keshawa,Morozov,haase,Kempers,Wurger,Dougherty,Morteza}, the Soret coefficient
is related to the macroscopic thermodynamic quantities such as the
heat of transport and the molar enthalpy.

{
Molecular dynamics (MD) simulations are useful to investigate the Soret coefficients of liquids directly\cite{kohler,artola}.  
In MD simulations, roughly there are two different methods to calculate the Soret coefficient. 
One is the non-equilibrium MD (NEMD) simulation method, in which a temperature gradient is applied to the system. 
Non-equilibrium physical quantities such as the heat flow can be directly calculated in a NEMD simulation,
and the Soret coefficient can be estimated without phenomenological assumptions.
To calculate the heat flow efficiently, we can utilize the reverse NEMD (RNEMD) method\cite{Reith,Galliero1,Galliero2}. 
In the RNEMD method, the heat flow is imposed to the system and the temperature gradient is measured. 
With this method, we can avoid the statistically inefficient calculation of the heat flow. 
Reith and M\"{u}ller-Plathe \cite{Reith}, and Galliero et al \cite{Galliero1,Galliero2} utilized the RNEMD method to calculate the Soret coefficient of binary LJ liquids.
}
Another is the equilibrium MD (EMD) method.
In this approach, the Soret coefficient is obtained from the linear response theory in equilibrium\cite{D.todd}.
According to the linear response theory, transport coefficients can generally be calculated from a correlation function of the fluctuations of the flux at equilibrium.
For instance, Sarman and Evans\cite{Sarman} reported that the EMD reproduces the consistent results with NEMD.
Hoheisel and Vogelsang\cite{Vogelsang1988} conducted a systematic study to report that particles with larger mass and larger cohesive energy move to the cold side.
Vogelsang et al.\cite{Vogelsang1987} introduced an interesting analysis method which utilizes the decomposition of fluxes.
They decomposed the heat flux into kinetic, potential and enthalpy contributions to report 
that the enthalpy contribution is dominant in the thermal diffusion coefficient (and thus in the Soret coefficient).
However, they presented the result only for a specific interaction that mimics \ce{Ar}-\ce{Kr} mixtures.

{
Motivated by the work by Vogelsang et al, in this work we studied
the contributions of different heat flux components to the transport
coefficient. We employed equimolar binary liquid mixtures which have
the constant mass ratio and similar liquid structures but different
interaction potentials. By changing the inter-particle interaction
systematically, the dynamic properties can be changed while the static
liquid structure is almost unchanged. We decomposed the heat flux
into the kinetic and potential contributions, and calculated the
cross-correlation functions for the mass flux and the decomposed heat
fluxes. Based on the cross-correlation functions, we discuss which
contribution is sensitive to the inter-particle potential.
Details are shown below. 
}
\section{Model and Analysis}
To provide a strategy of analysis, before presenting simulation details, 
let us consider a macroscopic non-equilibrium three-dimensional system where several fluxes are induced by thermodynamic forces. 
To discriminate the different fluxes, we use the subscript $i$ and express the $i$-th macroscopic flux as $\bm{J}_i$. We
express the thermodynamic force, which is conjugate to the $i$-th flux
as $-\bm{X}_i$.
The fluxes at a given position $\bm{r}$ and a given time $t$ are generally given as functionals of the thermodynamic force as

\begin{equation}
 \bm{J}_i(\bm{r},t)=\bm{J}_{i}[\bm{X}_i(\bm{r},t),\bm{r},t].\label{J} 
\end{equation}
If the system is near equilibrium and the spatial and temporal variation of the fluxes are broad and slow, Eq.~\eqref{J} can be phenomenologically rewritten as
\begin{equation}
 \bm{J}_i(\bm{r},t)=-\sum_{k}L_{ik}\bm{{X}}_k(\bm{r},t),\label{JL}
\end{equation}
where $L_{ik}$ are Onsager coefficients, which are second-rank polar tensor.

To study the Soret effect, we consider binary mixtures of fluids.
{As the fluxes, we consider the mass flux of particles 1 $\bm{J}_1(\bm{r},t)$, the mass flux of particles 2 $\bm{J}_2(\bm{r},t)$ and the heat flux $\bm{J}_{q}(\bm{r},t)$.
To satisfy the momentum conservation, $\bm{J}_1(\bm{r},t)+\bm{J}_2(\bm{r},t)=\bm{0}$. }
Then Eq.~\eqref{JL} can be rewritten as the following set of equations:

\begin{align}
 \bm{J}_{q}(\bm{r},t)&=-L_{qq}\bm{X}_{q}(\bm{r},t)-L_{q1}\bm{X}_{1}(\bm{r},t),\label{heat_L}\\                                                                                                                                 
 \bm{J}_{1}(\bm{r},t)&=-L_{11}\bm{X}_{1}(\bm{r},t)-L_{1q}\bm{X}_{q}(\bm{r},t).                                                                                                                                            
\end{align}
From the viewpoint of the transport phenomena, the mass and heat fluxes can be expressed phenomenologically in terms of the gradients of the temperature field and the volume fraction field\cite{D.todd}.

\begin{align}
 \bm{J}_{q}(\bm{r},t)&=-\lambda\nabla T-\rho\frac{\partial \mu_{1}}{\partial c_{1}}TD{''}\nabla c_{1},\label{heat_J}\\                                                                                     
 \bm{J}_{1}(\bm{r},t)&=-\rho D\nabla c_{1}-\rho c_{1}c_{2}D'\nabla T,                                                                                                                                     
\end{align}
where $\nabla T$ is the gradient of temperature, $\nabla c_1$ is the gradient of volume fraction of particle 1, $D''$ is the Dufour coefficient, $\lambda$ is the thermal conductivity, 
$\rho$ is the mass density, and $\mu_1$ is the chemical potential of particle 1. 
By comparing Eqs.~\eqref{heat_L} and \eqref{heat_J}, $D$ and $D'$ can be written as\cite{D.todd}
\begin{align}
 D&=\frac{L_{11}}{\rho c_{2}T}\left(\frac{\partial \mu_{1}}{\partial c_{1}}\right),\\
 D'&=\frac{L_{1q}}{\rho c_{1}c_{2}T^2}.
\end{align}
The Soret coefficient $S_T$ can be obtained from $D$ and $D'$ according to Eq.~\eqref{Soret_coefficient}.
To calculate $S_T$, we need to obtain $\partial \mu_1 / \partial c_1$. Acquisition of chemical potential by EMD is generally difficult. 
In this work, we discuss $L_{1q}$ instead of the Soret effect. 
Indeed, $L_{1q}$ has the same sign as $S_T$ (as far as $\partial \mu_1 / \partial c_1$ is positive), and can be calculated accurately.
We limit ourselves to simple systems where $\partial \mu_1 / \partial c_1>0$.

Meanwhile, in our EMD simulations, we consider equimolar binary mixtures of fluids, for which the mass of the particle 1 is $m$ and the mass of the particle 2 is $M$.
The inter-particle interaction $U(r)$ is common for all the particle pairs in the system, and it is written as follows:
\begin{equation}
   U(r)=\begin{cases}
     4\varepsilon\left[\left(\frac{\sigma}{r}\right)^{12}-\left(\frac{\sigma}{r}\right)^6\right] +U_c&(r\leq r_c),\\ 
     0 &(r> r_c).
        \end{cases}
\end{equation}
Here, $r$ is the distance between two particles, $\sigma$ is the particle size and $\varepsilon$ is the intensity parameter.
$U_c$ is the potential shift to attain $U(r_c)=0$ at $r=r_c$. 
{We chose units of length, energy and mass as $\sigma$, $\varepsilon$, and $m$.}
Namely, for the case with $r_c=3.0$, we have an attractive part, whereas with $r_c=2^{1/6}\sigma$ the interaction is purely repulsive. 
The static liquid structure, which can be characterized by the radial distribution function, was not sensitive to the value of $r_c$ in the examined parameter range.
{We performed the EMD simulations with $32000$ particles with the density at $0.7$, the mixing ratio at $0.5$, the mass ratio $M$ at $2.0$}, and the normalized temperature defined as $k_BT/\varepsilon$ at $1.5$ { to minimize the fluctuations of macroscopic fluxes\cite{Smit}. }
The equations of motion were integrated with the velocity Verlet algorithm\cite{Computer} in LAMMPS\cite{Lammps}, and the integration step size was $\delta t=0.0005$.
$\sigma$ and $\varepsilon$ were fixed at unity. 
We performed simulations with $1.0\times 10^8$ steps ($t=5.0\times 10^4$).
Before the data acquisition, we set the temperature as $1.5$ by using Nos\'e-Hoover thermostat.
{To equilibrate the system sufficiently, we performed simulations with $1.0\times 10^7$ steps ($t=5.0\times 10^3$) before the data acquisition.}

{The Onsager coefficients can be calculated from the correlation functions calculated in the EMD simulations. 
The linear response theory\cite{D.todd} relates the correlation functions to the Onsager coefficients. 
We expect that the simulation box is much smaller than the characteristic length scale of the macroscopic fields. 
Then the Onsager coefficients are calculated by using the correlation function of the integrated total fluxes in a simulation box.
The mass flux of the MD system $\hat{\bm{J}}_1(t)$ is expressed as follows, in terms of the microscopic state:
}
\begin{equation}
  \hat{\bm{J}}_1(t)=\sum_{i=1}^{N_1}m_i\bm{v}_i(t) \label{mass_flux},  
\end{equation}
where  $m_i$ is the mass of the $i$-th particle, $N_1$ is the number of particle 1 and $\bm{v}_i(t)$ is the velocity of the $i$-th particle at time $t$.
However, the microscopic heat flux is not simple and there are different expressions for the microscopic heat flux.
In this study, we employ the Irving and Kirkwood\cite{Irving1950} expression for $\hat{\bm{J}}_q(t)$ by interpreting the heat flux as the energy flux:
\begin{align}
 \hat{\bm{J}}_q(t)&=\sum_{i=1}^{N}\left[\frac{1}{2}m_i\bm{v}_i^2(t)+\sum_{j\neq i}U_{ij}(r_{ij}(t))\right]\bm{v}_i(t)-\frac{1}{2}\sum_{i=1}^{N}\sum_{j\neq i}\bm{r}_{ij}(t)\frac{\partial U_{ij}(r_{ij}(t))}{\partial \bm{r}_{ij}(t)}\cdot\bm{v}_i(t)\label{heat_flux},
\end{align}
where $\bm{r}_{i}(t)$ is the position of the $i$-th particle at time $t$, $\bm{r}_{ij}(t) = \bm{r}_{i}(t) - \bm{r}_{j}(t)$ and $N$ is the total number of particles.
The Onsager coefficients $L_{11}$ and $L_{1q}$ can be expressed in terms of the equilibrium correlation functions as follows\cite{D.todd}:
{
\begin{align}
 L_{11}=\frac{1}{3Vk_B}\int_{0}^{\infty}dt\langle \hat{\bm{J}}_{1}(t)\cdot\hat{\bm{J}}_{1}(0)\rangle,\label{autocorrelation_function}\\                                               
 L_{1q}=\frac{1}{3Vk_B}\int_{0}^{\infty}dt\langle \hat{\bm{J}}_{1}(t)\cdot\hat{\bm{J}}_{q}(0)\rangle.\label{crosscorrelation_function}                                                     
\end{align}
Here, $\langle ... \rangle$ represents the equilibrium
statistical average, $V$ is the volume. 
Note that $\hat{\bm{J}}_{1}(t)$ and $\hat{\bm{J}}_{q}(t)$
are microscopically defined and they fluctuates with time.
{We multiply the factor of $1/3$ because the system is isotropic.   
This integral is evaluated by the trapezoidal rule after equilibrium molecular dynamics simulations. }
\section{Results}
{
Figures~\ref{L11} and \ref{L1q} show $L_{11}$ and $L_{1q}$ as functions of the cut-off length $r_c$ of the potential. 
Note that the magnitude of statistical error is smaller than the symbol in the plots.
Figure~\ref{L11} demonstrates that $L_{11}$ decreases with increasing $r_c$ at $r_c<1.8$,
it shows a minimum around $r_c\approx 1.8$ and approaches to a steady value in $r_c>2.5$. 
$L_{1q}$ shown in Fig.~\ref{L1q} also decreases with increasing $r_c$. 
However, it monotonically decreases without showing any minima within the examined range. 
Further, $L_{1q}$ goes down to negative around $r_c\approx 2.0$. 
This change in $L_{1q}$ corresponds to the change of sign for $S_T$. 
As we stated, the positive $L_{1q}$ for small $r_c$ means that the lighter particles migrate toward the low-temperature side. 
When the sign of $S_T$ changes with increasing $r_c$, the lighter particles move to the opposite direction.
\begin{figure}[H]
    \centering
   \includegraphics[width=72mm]{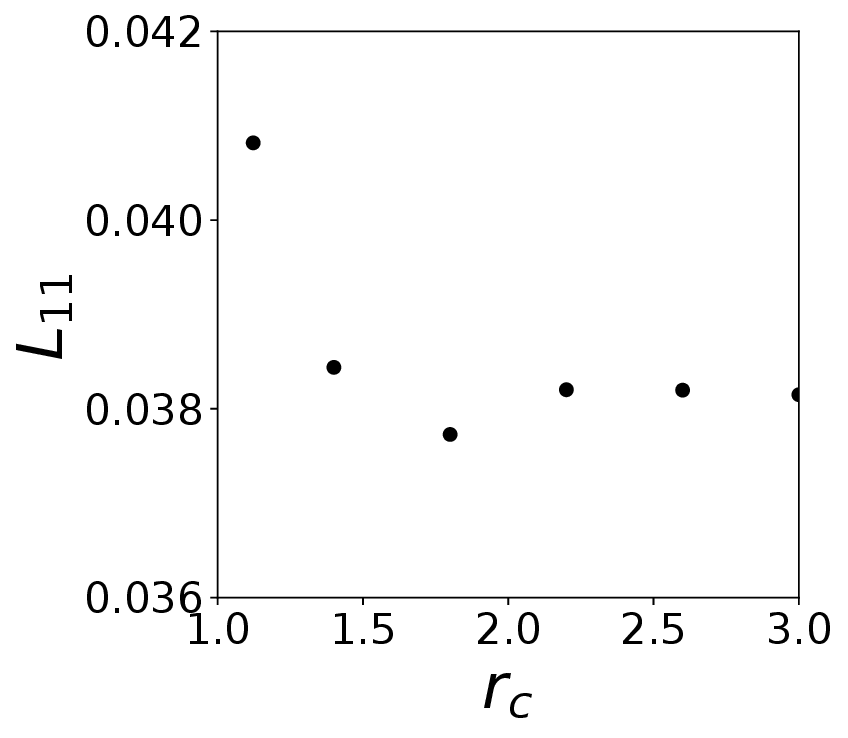}
 \caption{$L_{11}$ as a function of $r_c$.}\label{L11}
\end{figure}
\begin{figure}[H]
   \centering
   \includegraphics[width=72mm]{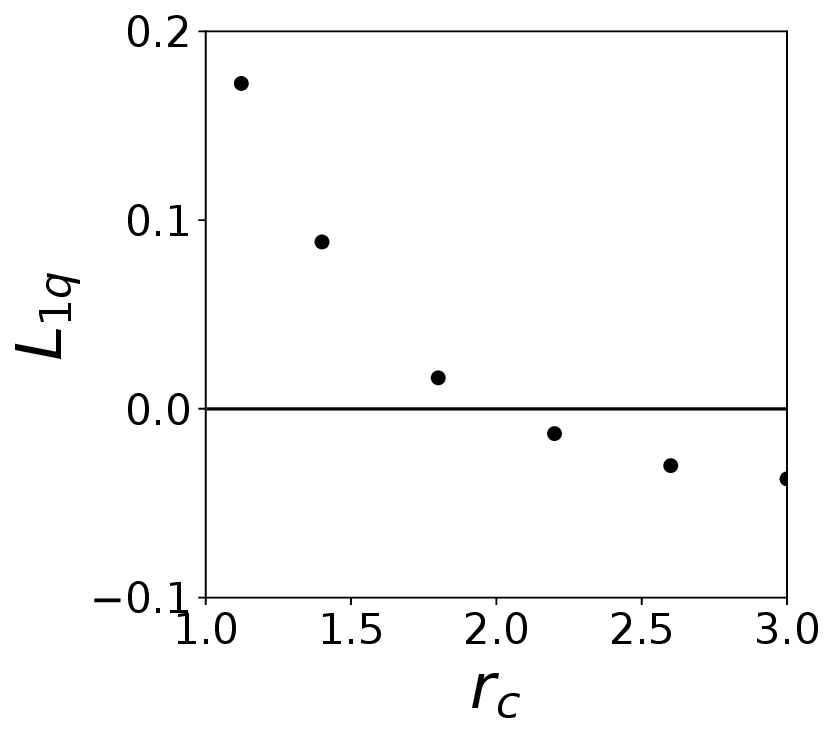}
  \caption{$L_{1q}$ as a function of $r_c$. Horizontal line indicates $L_{1q}=0$.}\label{L1q}
\end{figure}
To analyze the change of $L_{1q}$ induced by $r_c$, 
we observe the cross-correlation function $C_{1q}$ defined as $\left\langle\hat{\bm{J}}_1(t)\cdot\hat{\bm{J}}_{q}(0)\right\rangle$. 
Figure~\ref{C1q} shows $C_{1q}$ for various $r_c$. 
In the case of small $r_c$, $C_{1q}$ monotonically decays with time. 
As $r_c$ increases, $C_{1q}$ gradually exhibits an undershoot, and the magnitude of undershoot increases.
As a result of this undershoot, $C_{1q}$ becomes negative, and it causes a negative $L_{1q}$ in Fig.~\ref{L1q}. 
\begin{figure}[H]
 \begin{center}
 \includegraphics[width=72mm]{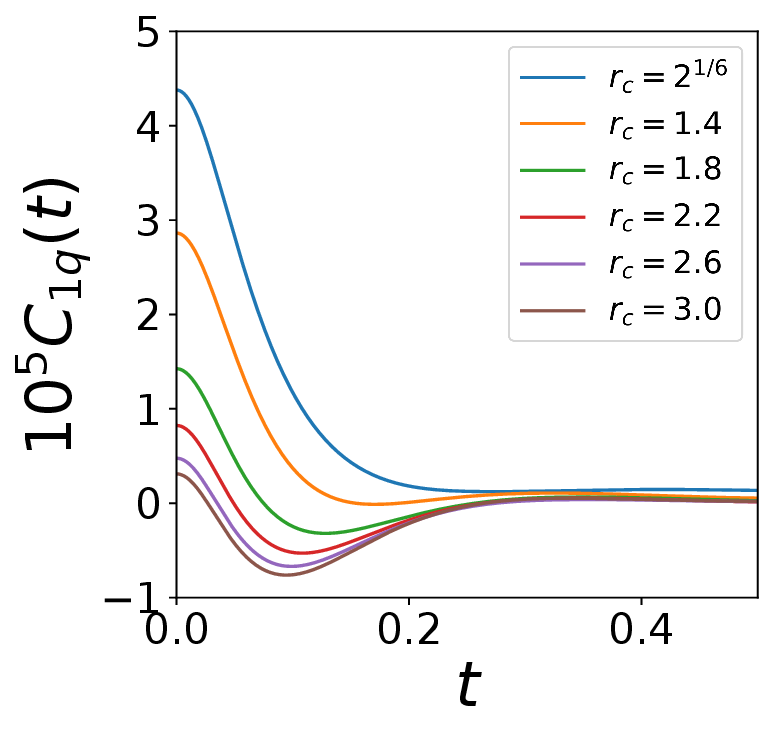}
  \end{center}
 \caption{Cross-correlation functions $C_{1q}$ for various $r_c$ values.}\label{C1q}
 \end{figure}

To see the origin of the undershoot in $C_{1q}$,} we decompose the energy flux $\hat{\bm{J}}_{q}(t)$ into the kinetic and potential contributions 
($\hat{\bm{J}}^{(K)}_q(t)$ and $\hat{\bm{J}}^{(P)}_q(t)$, respectively) as shown below:
\begin{align}
 \hat{\bm{J}}_q(t)= \hat{\bm{J}}^{(K)}_q(t)+ \hat{\bm{J}}^{(P)}_q(t)\label{composed_heat_flux} ,
\end{align}
\begin{equation}
 \hat{\bm{J}}^{(K)}_q(t)=\sum_{i=1}^{N}\left[\frac{1}{2}m_i\bm{v}_i^2(t)\right]\bm{v}_i(t)\label{heat_kinetic}, 
\end{equation}
{
\begin{equation}
 \hat{\bm{J}}^{(P)}_q(t)=\sum_{i=1}^{N}\left[\sum_{j\neq i}U_{ij}(r_{ij})\bm{v}_i(t)-\frac{1}{2}\sum_{i=1}^{N}\sum_{j\neq i}\bm{r}_{ij}\frac{\partial U_{ij}(r_{ij})}{\partial \bm{r}_{ij}}\cdot\bm{v}_i(t)\right]\label{heat_potential}.
\end{equation}
}
For the flux given in Eqs.~\eqref{heat_kinetic} and \eqref{heat_potential}, we calculated the cross-correlation functions 
$C_{1q}^{(K)}=\left\langle \hat{\bm{J}}_1(t)\cdot\hat{\bm{J}}_q^{(K)}(0)\right\rangle$ and $C_{1q}^{(P)}=\left\langle \hat{\bm{J}}_1(t)\cdot\hat{\bm{J}}_q^{(P)}(0)\right\rangle$
\begin{figure}[H]
 \begin{tabular}{cc}
  \begin{minipage}{0.45\hsize}
   \centering
   \includegraphics[width=60mm]{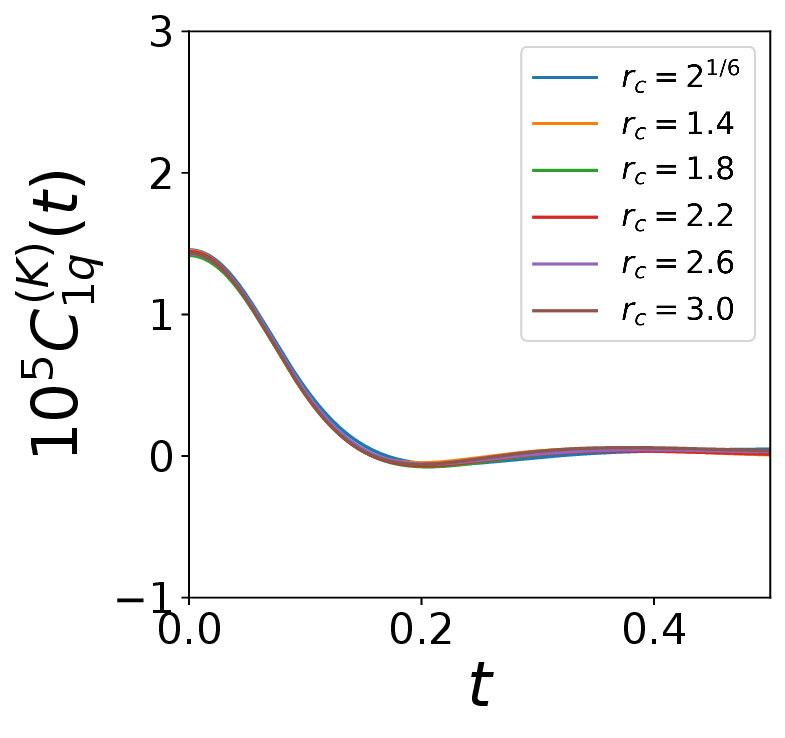}
  \end{minipage}
  \begin{minipage}{0.45\hsize}
   \centering
   \includegraphics[width=60mm]{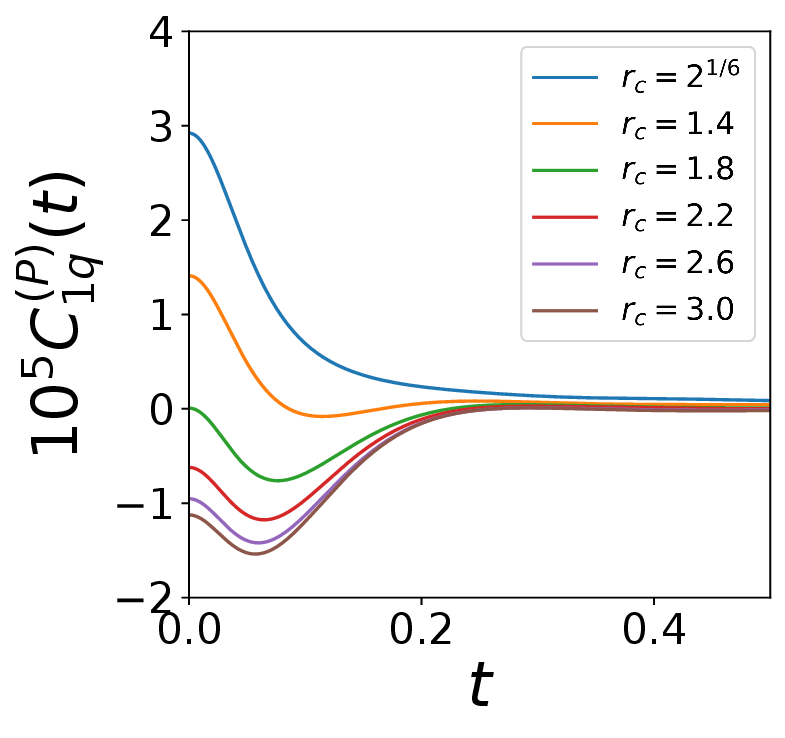}
  \end{minipage}
 \end{tabular}
\caption{Cross-correlation functions $C_{1q}^{(K)}$ and $C_{1q}^{(P)}$ for various $r_c$ values.}\label{C1q_Compose}
\end{figure}
{
Figure~\ref{C1q_Compose} shows the cross-correlation functions thus calculated, demonstrating that the kinetic contribution is not sensitive to $r_c$. 
{Here, we should stress that the insensitivity of $C_{1q}^{(K)}(t)$ to $r_c$ is not trivial. 
As we explained, the fluid structure is not sensitive to $r_c$, but the dynamical properties such as the diffusion coefficient generally depend on $r_c$.
}
In contrast, the potential contribution is significantly dependent on $r_c$. 
These results clearly demonstrate that the interaction dependence of $L_{1q}$ shown in Fig.~\ref{L1q} comes from $C_{1q}^{(P)}$. 
In particular, the negative $L_{1q}$ is due to the undershoot in $C_{1q}^{(P)}$ in $t<0.2$. 

\section{DISCUSSIONS}
We note that the results may be altered if a different heat flux expression is employed.
{For instance, Hoheisel and Vogelsang \cite{Vogelsang1988} employed the Bearman-Kirkwood expression\cite{Macgowan1986,onuki} shown below:}
\begin{equation}
 \hat{{\bm{J}}}^{Bearman}_q(t)= \hat{\bm{J}}_q(t)-\sum_{\nu}^{2}\sum_{i}^{N_i}h_{\nu}\frac{\bm{v}_{\nu i}}{m_{\nu}},\label{bearman}
\end{equation}
where $h_{\nu}$ is the molar enthalpy of particles $\nu$.
Using this heat flux, they obtained $L_{1q}$ for a soft-core system and reported that the sign of $L_{1q}$ is negative. 
This result contradicts our results in Fig.~\ref{L1q} at $r_c=2^{1/6}$. 
Namely, Eq.~\eqref{bearman} employs $h_{\nu}$, which is not straightforwardly obtained from microscopic simulations.
We note that temperature and potential are also different from ours, although the effect seems not significant.
This inconsistency is due to the expression of the heat flux.

We also note that Sasa\cite{sasa} has recently proposed the other expression of the heat flux as written below:
\begin{equation}
 \hat{\bm{J}}_q^{Sasa} (t)\equiv \int_V d\bm{r}\left(\hat{\bm{J}}_q^{\prime}(\bm{r},t)-\frac{e^{\prime}(\bm{r},t)+p(\bm{r},t)}{\rho(\bm{r},t)} \hat{\bm{\pi}}^{\prime}(\bm{r},t)\right),\label{sasa}
\end{equation}
where $\hat{\bm{J}}_q^{\prime}$, $e^{\prime}$ and $\hat{\bm{\pi}}^{\prime}$, are the energy density flux, the energy density and the momentum density in the moving frame with the local velocity. 
These quantities should be defined in the moving frame since the contribution of the local velocity should be subtracted.  
$\rho(\bm{r},t) $ is the mass density and $p(\bm{r},t)$ is the thermodynamic pressure derived from the local thermodynamic entropy.
Eq.\eqref{sasa} has a rigid microscopic origin, obtained from the Hamiltonian dynamics and local equilibrium assumptions for a chosen coarse-grained length scale. 
However, this heat flux is not easily obtained from the trajectories in EMD simulations.

We also note that Sasa theory includes a physical quantity corresponding to the molar enthalpy like Bearman-Kirkwood.
Therefore, in order to correctly describe the macroscopic transport, the contribution of molar enthalpy must also be taken into account. 
However, even if the molar enthalpy contribution is added, the expressions for the kinetic energy and potential energy contributions for the heat flux remains the same.
If the correct heat flux with the enthalpy contribution is employed, we have the third cross-correlation function for the enthalpy flux.
The difference between our results and that by Hoheisel and Vogelsang\cite{Vogelsang1988} means that the enthalpy flux has so strong a contribution that the sign of the transport coefficient is changed. 
As future studies, it is demanding to obtain the cross-correlation function for the enthalpy flux in our system.
Then we will be able to study how different correlation functions affect the transport coefficient and which one is dominant.

\section{Conclusions}
To see the contributions of kinetic and potential origins in the heat flux to the transport coefficient $L_{1q}$, 
we conducted molecular dynamics simulations for equimolar binary fluids with the inter-particle potential of various cut-off lengths $r_c$. 
The simulation revealed that $L_{1q}$ changes its sign by $r_c$. 
Specifically, $L_{1q}<0$ for the case with $r_c>2$, whereas $L_{1q}>0$ for small $r_c$.
}
We decomposed the heat flux into the kinetic and potential contributions, 
and calculated the cross-correlation functions between the mass flux and the decomposed heat fluxes. 
The result demonstrated that the sign of $L_{1q}$ is dominated by the potential contribution. 

It would be fair to mention that our results may be changed
if other expressions for the heat flux are employed instead of
the Irving-Kirkwood heat flux. 
Although the Irving-Kirkwood heat flux is not thermodynamically correct,
our cross-correlation functions for the kinetic and potential energy
fluxes are correct. We expect that how these cross-correlation functions
depend on the cutoff provides useful information to discuss the
cross-transport coefficient with the enthalpy contribution.
The calculations of $L_{1q}$ with different
expressions of the heat flux including the statistical mechanically
correct expression by Sasa\cite{sasa} will be required. It would be an
interesting and important future work.

\section{Acknowledgment}
The authors thank Prof. Sasa (Kyoto University) for informing his work on the derivation of hydrodynamic equations from the Hamiltonian dynamics.




\begin{thebibliography}{99}
\bibitem{platten}J. K. Platten, Journal of Applied Mechanics \textbf{73}, 5 (2006).
  \bibitem{chapman}S. Chapman, and T. G. Cowling, The Mathematical Theory of Non-uniform Gases (Cambridge University Press, Cambridge, 1970).
 \bibitem{Dootson}S. Chapman and F. W. Dootson, Philos. Mag. \textbf{33}, 248 (1917).
 \bibitem{bluh}G. Bl\"{u}h, O. Bl\"{u}h, and M. Puschner, Philos. Mag. 24, 1103 (1937).
 \bibitem{Ibbs}T. S. Ibbs, Roy. Soc. Proc. \textbf{93}, 148 (1916)
 \bibitem{snowdon1960}P. N. Snowdon and J. C. R. Turner, Trans. Faraday Soc. \textbf{56}, 1812 (1960).
 \bibitem{Tanner}C. C. Tanner, Trans. Faraday Soc. \textbf{23}, 75 (1927).
 \bibitem{Tanner1}C. C. Tanner, Trans. Faraday Soc. \textbf{49}, 611 (1953).
 \bibitem{Prigogine.1952}I.Prigogine, L. Brouckere, and R. Buess, Physica \textbf{18}, 915 (1952).
 \bibitem{Prigogine.1950}I.Prigogine, L. Brouckere, and R. Amand, Physica \textbf{16}, 851 (1950).
 \bibitem{Saxton}R. L. Saxton, E. L. Dougherty, and H. G. Drickamer, J. Chem. Phys. \textbf{22}, 1166 (1954).
 \bibitem{korsching}H. Korsching, Naturwissenschaften \textbf{31}, 348 (1943).
 \bibitem{wirtz}K. Wirtz, Z. Naturforsch \textbf{3a}, 672 (1948).
\bibitem{Denbigh}K. G. Denbigh, Trans. Faraday Soc. \textbf{48}, 1 (1952).
\bibitem{Hartmann}S. Hartmann, G. Wittko, F. Schock, W. Grob, W. K. F. Lindner, and K. I. Morozov, J. Chem. Phys. \textbf{141}, 134503 (2014).
\bibitem{Hartmann2}S. Hartmann, G. Wittko, and W. Kohler, Phys. Rev. Lett. \textbf{109}, 65901 (2012).
\bibitem{Keshawa}K. Shukla and A. Firoozabadi, Ind. Eng. Chem. Res. \textbf{37}, 3331 (1998).
\bibitem{Morozov}I. Morozov, Phys. Rev. E \textbf{79}, 31204 (2009).
\bibitem{haase}R. Haase, Zeitschrift f\"{u}r Physik \textbf{127}, 1 (1950).
\bibitem{Kempers}L. J. T. M. Kempers, J. Chem. Phys. \textbf{115}, 6330 (2001).
\bibitem{Wurger}A. Wurger, J. Phys. Condens. Matter \textbf{26}, 35105 (2014).
\bibitem{Dougherty}E. L. Dougherty and H. G. Drickamer, J. Chem. Phys. \textbf{23}, 295 (1955).
\bibitem{Morteza} M. Eslamian and M. Z. Saghir, Phys. Rev. E \textbf{80}, 11201 (2009).
\bibitem{kohler}W. Kohler and S.Wiegand, Thermal Nonequilibrium Phenomena in Fluid Mixtures (Springer, 2008).
\bibitem{artola}P. A. Artola and B. Rousseau, Mol. Phys. \textbf{111}, 3394 (2013).
\bibitem{Reith}D. Reith, and F. M\"{u}ller-Plathe, J. Chem. Phys., \textbf{112}, 2436 (2000).
\bibitem{Galliero1}G. Galliero, B. Duguay, J. P. Caltagirone, and F. Montel, Philos. Mag. \textbf{83}, 2097 (2003).
\bibitem{Galliero2}G. Galliero, B. Duguay, J. P. Caltagirone, and F. Montel, Fluid Phase Equilib. \textbf{208}, 171 (2003).
 \bibitem{Sarman}S. Sarman and D. J. Evans, Phys. Rev. A \textbf{45}, 2370 (1992).
\bibitem{Vogelsang1988}C. Hoheisel, R. Vogelsang, J. Chem. Phys. \textbf{89}, 174503 (1988).
\bibitem{Vogelsang1987}R. Vogelsang, C. Hoheisel, G. V. Paolini and G. Ciccotti, Phys. Rev. A \textbf{36}, 3964 (1987).
\bibitem{D.todd}B. D. Todd and P. J. Daivis, Nonequilibrium Molecular Dynamics (Cambridge University Press, Cambridge, 2017).
\bibitem{Smit}B. Smit, J. Chem. Phys. \textbf{96}, 8639 (1992).
\bibitem{Computer}M. P. Allen and D. J. Tildesley, Computer Simulation of Liquids (Oxford University Press, Oxford, 1986).
\bibitem{Lammps}S. Plimpton, J. Comput. Phys. \textbf{1}, 117 (1995).
\bibitem{Irving1950}J. H. Irving and J. G. Kirkwood, J. Chem. Phys. \textbf{18}, 338 (1950).
\bibitem{Macgowan1986}D. MacGowan and D. J. Evans, Phys. Rev. A \textbf{34}, 2113 (1986).
\bibitem{onuki}A. Onuki, J. Chem. Phys. \textbf{151}, 134118 (2019).
\bibitem{sasa}S. Sasa, Phys. Rev. Lett. \textbf{112}, 100602 (2014).
\end{thebibliography}

\end{document}